\documentclass[review,3p,twocolumn]{elsarticle}

\journal{Physics Letters A}
\biboptions{sort&compress}

\usepackage{amsmath}
\usepackage{amssymb}
\usepackage[T1]{fontenc}
\DeclareUnicodeCharacter{03B1}{\ensuremath{\alpha}}
\DeclareUnicodeCharacter{2212}{\ensuremath{-}}
\DeclareUnicodeCharacter{00A0}{~}
\DeclareUnicodeCharacter{2013}{--}
\DeclareUnicodeCharacter{2014}{---}
\DeclareUnicodeCharacter{2192}{\ensuremath{\rightarrow}}
\graphicspath{{pictures/}{./pictures/}}
\usepackage{graphicx}
\usepackage{bm}
\usepackage{xcolor}
\usepackage{tikz}
\usepackage{float}
\usepackage{subcaption}
\usepackage{pgfplots}
\usepackage{enumitem}
\usepgfplotslibrary{groupplots}
\pgfplotsset{compat=1.18}
\usetikzlibrary{calc,arrows.meta,patterns}
\usepackage[colorlinks=true, citecolor=blue, linkcolor=red, urlcolor=blue]{hyperref}
\usepackage{lineno}

\begin{document}
\begin{frontmatter}

\title{Staggered Potential and Elliptical Light Driven Topological Phase Transitions in $\alpha$-$\mathcal{T}_{3}$ Lattice }

\author{Muhammad Faisal}
\ead{faisal.physics.qau@gmail.com}

\author{Muhammad Irfan Sarwar\corref{cor1}}
\ead{isarwar@phys.qau.edu.pk}

\address{Department of Physics, Quaid-I-Azam University, Islamabad 45320, Pakistan}

\cortext[cor1]{Corresponding author.}

\begin{abstract}
We theoretically investigate the influence of hexagonal boron nitride
(h-BN) on the electronic properties of an $\alpha$-$\mathcal{T}_{3}$
lattice driven by an off-resonant elliptically polarized light field.
The staggered potential $M$ breaks the sublattice inversion symmetry,
transforming the initial semimetal into a trivial insulator with Chern
number $C=0$.
We identify a fundamental geometric singularity at $\alpha=1/\sqrt{2}$,
independent of $M$, where the lower gap-closing threshold diverges.
Because the elliptical drive breaks time-reversal symmetry, this
threshold is valley-resolved, and together with the upper threshold it
bounds a finite topological window where conduction--flat band inversion
yields a Chern insulator with $C=1$ carried by the flat band.
Above this critical value the lower gap closes at a finite intensity,
allowing a transition from $C=1$ to $C=2$ as the dice limit ($\alpha=1$)
is approached.
The topological phases are characterized by quantized anomalous Hall
plateaus at $\sigma_{xy}=e^2/h$ ($C=1$) and $\sigma_{xy}=2e^2/h$
($C=2$), with each valley contributing exactly $\frac{1}{2}e^2/h$.
The robustness of each plateau is set by the gap in which the Fermi
level lies: the flat-band $C=1$ plateau sits in a narrow gap and is most
fragile, while the $C=2$ plateau is protected by a wider gap and remains
robust to room temperature.
A highly asymmetric thermoelectric Seebeck response further serves as
an experimental fingerprint of each phase, providing a realistic
framework for realizing stable high-Chern-number phases in
substrate-supported $\alpha$-$\mathcal{T}_{3}$ materials.
\end{abstract}


\end{frontmatter}

%%%===================================================================
\section{Introduction}
%%%===================================================================

Condensed matter physics was transformed from the era of the classical
Hall effect by the discovery of the Quantum Hall Effect
(QHE)~\cite{stone1992quantum,prange1990quantum} and Topological
Insulators (TIs)~\cite{rudner2019floquet,hasan2010colloquium,fu2007topological}.
These phases are characterized not by local order parameters but by
topological invariants: the $Z_2$ invariant~\cite{moore2007topological}
and the Chern number~\cite{hatsugai1993chern,thouless1982quantized}.   
Chern insulators and topological phase transitions have recently emerged as a central topic in condensed-matter physics, owing to their novel quantum phenomena and promising technological
applications~\cite{PhysRevLett.136.046602,2025ACSNa..1935575Z, acsnano.5c00323,PhysRevB.112.085128,https://doi.org/10.1002/adfm.202501934}.
While the conventional Haldane model~\cite{jotzu2014experimental}
achieves topological band structure through complex hopping phases,
Floquet theory provides a more general and experimentally accessible
route to the same physics~\cite{eckardt2017colloquium}.
Topological phases can be induced in a material by periodic driving
fields such as a laser~\cite{eckardt2017colloquium}.
These Floquet Topological Insulators
(FTIs)~\cite{cayssol2013floquet,rudner2019floquet} offer a significant
advantage over static TIs: their topological properties can be
switched on and off on femtosecond timescales by controlling the
external field~\cite{cayssol2013floquet,li2021phase}.

Among all proposed FTI platforms, two-dimensional Dirac
materials~\cite{wehling2014dirac} have received the greatest
attention.
The opening of a Floquet-induced gap in graphene was first predicted
and experimentally observed~\cite{wang2013observation}, arising from
the linearly dispersing Dirac fermions at the valleys $K$ and
$K'$~\cite{jiang2007quantum,berkolaiko2018symmetry}.
Recent research has shifted toward the $\alpha$-$\mathcal{T}_3$
lattice, which interpolates continuously between the honeycomb lattice
of graphene ($\alpha=0$) and the dice lattice ($\alpha=1$)
~\cite{illes2017properties,raoux2014orbital,vidal1998aharonov,wang2021quantum}.
The $\alpha$-$\mathcal{T}_3$ model is of particular interest because
it contains not only Dirac cones but also a non-dispersive flat band
(exact for $\alpha=1$), whose divergent density of states can amplify
correlation effects and give rise to topological
phases~\cite{illes2017properties,balassis2020magnetoplasmons,leykam2018artificial}.

\begin{figure}[!t]
    \centering
    \includegraphics[width=\linewidth]{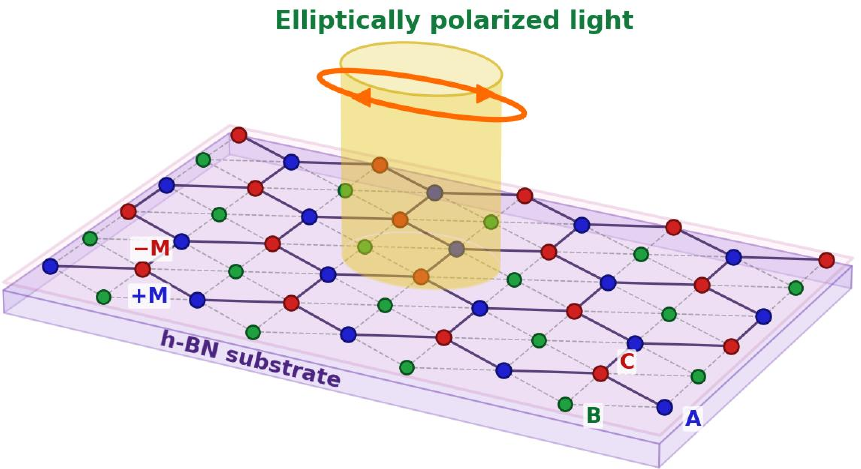}
    \caption{Schematic of the $\alpha$-$\mathcal{T}_3$ lattice on an
    h-BN substrate driven by an off-resonant elliptically polarized
    laser. Rim sites $A$ and $C$ acquire staggered energies $\pm M$
    while the hub site $B$ remains at zero.}
    \label{fig:lattice}
\end{figure}

A fundamental experimental challenge in the implementation of Floquet
theory~\cite{rodriguez2021low} is the unavoidable influence of the
substrate on the 2D material~\cite{dai2019strain}.
The standard approach to improving carrier mobility in 2D materials is
to place them on a hexagonal boron nitride (h-BN)
substrate~\cite{dai2019strain,roy2021structure,dean2010boron}.
The substrate interaction and twistronics-induced Moiré
patterns~\cite{he2021moire,vogl2023light} create a static crystal
potential that breaks sublattice symmetry,
opening a trivial ($C=0$) band gap~\cite{ren2020twistronics}.
This static mass $M$ induced by the substrate then interacts with the
dynamic Floquet mass $\Delta_F$ generated by the applied 
polarized laser field~\cite{iurov2024floquet}.
While the effects of circular polarization on the massless
$\alpha$-$\mathcal{T}_3$ lattice have been
studied~\cite{PhysRevB.99.205429}, the rich interplay between a
substrate-induced static mass and a Floquet drive, including its
thermoelectric signatures, remains largely unexplored.

In this work we address this gap by mapping the complete topological
phase diagram of the massive $\alpha$-$\mathcal{T}_3$ lattice under
elliptical Floquet driving.
We study the realistic case in which the $\alpha$-$\mathcal{T}_3$
lattice is subject to both a staggered static
potential~\cite{sur2021floquet} and an off-resonant elliptical laser
field.
We show that the topological properties of the lattice are not
destroyed by the substrate; rather, there exists a critical phase
boundary where the dynamic Floquet interaction overcomes the static
mass, inverting the band gap and restoring a Chern insulator.
In the absence of the staggered potential, the lattice undergoes a
transition from $C=1$ to $C=2$ at
$\alpha=1/\sqrt{2}$~\cite{PhysRevB.99.205429}.
Interestingly, we find that at $\alpha=1/\sqrt{2}$ there exists a
geometric singularity (Fig.~\ref{fig:PhaseDiagram}) in the presence
of the staggered potential: the radiation intensity required to close
the gap between the valence and flat bands diverges at this point.
Our analysis reveals that the valence and flat bands never close for
$\alpha \leq 1/\sqrt{2}$ (Fig.~\ref{fig:BandEvolution}(a--c)), while
above this critical value the lower gap-closing intensity decreases
and both gaps close simultaneously only in the dice lattice limit
($\alpha=1$).

The paper is organized as follows.
In Sec.~\ref{sec:theory} we derive the static tight-binding
Hamiltonian with the h-BN staggered potential.
In Sec.~\ref{sec:trivial} we establish the topological baseline
of the undriven system and derive the effective Floquet Hamiltonian
using the high-frequency Floquet--Magnus
expansion~\cite{bukov2015universal}.
In Sec.~\ref{sec:gaps} we analytically derive the gap-closing
conditions and discuss the $\alpha=1/\sqrt{2}$ singularity.
In Sec.~\ref{sec:numerics} and Sec.~\ref{sec:phases} we present
numerical band structures showing the asymmetric phase evolution.
In Sec.~\ref{sec:berry} we characterize the topological phases
through Berry curvature distributions.
Sections~\ref{sec:hall} and~\ref{sec:thermo} present the quantized
anomalous Hall conductivity along with its temperature dependence
and thermoelectric transport, respectively. Finally, 
Conclusions are given in Sec.~\ref{sec:conclusions}.

%%%===================================================================
\section{Theoretical Framework}
\label{sec:theory}
%%%===================================================================

The total time-dependent Hamiltonian is the sum of the kinetic
hopping, the substrate-induced staggered mass, and the periodic drive:
\begin{equation}
 H(t) = H_{\mathrm{kin}} + H_{\mathrm{mass}} + H_{\mathrm{drive}}(t).
 \label{eq:total_H}
\end{equation}

\subsection{Static Hamiltonian with Staggered Potential}

We consider an $\alpha$-$\mathcal{T}_3$ lattice in the presence of a
staggered potential $M$ induced by a hexagonal boron nitride (h-BN)
substrate.
The tight-binding Hamiltonian in the basis
$\Psi_\mathbf{k} = (c_{A,\mathbf{k}},\, c_{B,\mathbf{k}},\, c_{C,\mathbf{k}})^T$
is given as:
\begin{gather}
 \hat{H}_{0}(\mathbf{k}) = \nonumber \\
 \begin{pmatrix}
  M & -\tau\cos\phi\, f(\mathbf{k}) & 0 \\
  -\tau\cos\phi\, f^*(\mathbf{k}) & 0 & -\tau\sin\phi\, f(\mathbf{k}) \\
  0 & -\tau\sin\phi\, f^*(\mathbf{k}) & -M
 \end{pmatrix},
 \label{eq:H0}
\end{gather}
where $\tau$ is the nearest-neighbor hopping amplitude and $\phi$
parametrizes the hub-site coupling via
\begin{equation}
 \alpha = \tan\phi, \qquad 0 \le \phi \le \frac{\pi}{4}.
 \label{eq:alpha_phi}
\end{equation}
The geometric structure factor is $f(\mathbf{k})=\sum_{j=1}^{3}
e^{i\mathbf{k}\cdot\boldsymbol{\delta}_j}$, where
$\boldsymbol{\delta}_j$ are the nearest-neighbor vectors~\cite{raoux2014orbital}.

At the Dirac points $K$ and $K'$, $f(\mathbf{k}_D)=0$
and the energy spectrum consists of three flat atomic levels:
$E=+M$ (conduction), $E=0$ (flat), and $E=-M$ (valence), creating a
trivial bandgap of $2M$ (Fig.~\ref{fig:static_band_structure}(a)).

Away from the Dirac points, the exact dispersion relation is determined by the
characteristic secular equation:
\begin{equation}
E^3 - \!\left(M^2 + \tau^2|f|^2\right)\!E
+ M\tau^2|f|^2\cos 2\phi = 0.
\label{eq:cubic}
\end{equation}
The final term in Eq.~(\ref{eq:cubic}) is proportional to $M\cos 2\phi$, explicitly breaks
particle-hole symmetry for all values of $\alpha$ except ($\alpha = 1$)
and produces a finite $\mathbf{k}$-dependent dispersion in the middle
band.
The symmetric spectrum
$E\in\{0,\pm\sqrt{M^2+\tau^2|f|^2}\}$ is recovered only in the
limiting cases $M=0$ (pristine graphene) or $\alpha=1$, $\phi=\pi/4$
(dice lattice).
This static, particle-hole-asymmetric configuration serves as the
trivial baseline from which all light-driven topological
transitions emerge.

%%%===================================================================
\section{Initial State: Trivial Insulator}
\label{sec:trivial}
%%%===================================================================

\begin{figure*}[htbp]
\centering
\begin{tikzpicture}

    % PANEL A: Flat atomic levels at the Dirac point
    \begin{scope}[xshift=0cm, scale=1.1]
        \draw[thick,->] (-0.5,-2.5) -- (-0.5,2.5) node[above] {$E$};
        \draw[dashed, gray] (-0.5,0) -- (2.5,0);
        \draw[ultra thick, blue]     (0,1.5)  -- (1.8,1.5)
            node[right, black] {$E_+ = +M$};
        \draw[ultra thick, darkgray] (0,0)    -- (1.8,0)
            node[right, black] {$E_0 = 0$};
        \draw[ultra thick, red]      (0,-1.5) -- (1.8,-1.5)
            node[right, black] {$E_- = -M$};
        \draw[<->, thick, darkgray] (0.8,-1.4) -- (0.8,1.4)
            node[midway, fill=white] {$2M$};
        \node[below] at (1,-2.5)
            {\textbf{(a)} At the Dirac point ($f(\mathbf{k})=0$)};
    \end{scope}

    % PANEL B: Dispersive bands away from the valley
    \begin{scope}[xshift=8.5cm, scale=1.1]
        \draw[thick,->] (-2.2,-2.5) -- (-2.2,2.5) node[above] {$E$};
        \draw[dashed, gray] (-2.2,0) -- (2.2,0);
        \draw[ultra thick, blue, domain=-1.8:1.8, samples=100]
            plot (\x, {sqrt(1.3+0.5*\x*\x)+0.05*\x*\x});
        \node[blue, right] at (1.8,2.0) {$E_+$};
        \draw[ultra thick, darkgray, domain=-1.8:1.8, samples=100]
            plot (\x, {0.22*\x*\x});
        \node[darkgray, right] at (1.8,0.8) {$E_0$};
        \draw[ultra thick, red, domain=-1.8:1.8, samples=100]
            plot (\x, {-sqrt(1.3+0.5*\x*\x)-0.05*\x*\x});
        \node[red, right] at (1.8,-2.0) {$E_-$};
        \filldraw[black] (0,1.14) circle (2pt);
        \filldraw[black] (0,0)    circle (2pt);
        \filldraw[black] (0,-1.14) circle (2pt);
        \draw[dashed, thin, gray] (0,-2.3) -- (0,2.3)
            node[above, black] {$K/K'$};
        \node[below] at (0,-2.5)
            {\textbf{(b)} Dispersive regime ($f(\mathbf{k})\neq0$)};
    \end{scope}

\end{tikzpicture}
\caption{Schematic band structure of the static substrate-supported
$\alpha$-$\mathcal{T}_3$ lattice ($\alpha<1$).
(a) Trivial band gap of $2M$ at Dirac points due to staggered potential $M$
(b) Away from the Dirac poins ($f(\mathbf{k})\neq0$), $M$ breaks
particle-hole symmetry and induces finite curvature in
bands.}
\label{fig:static_band_structure}
\end{figure*}
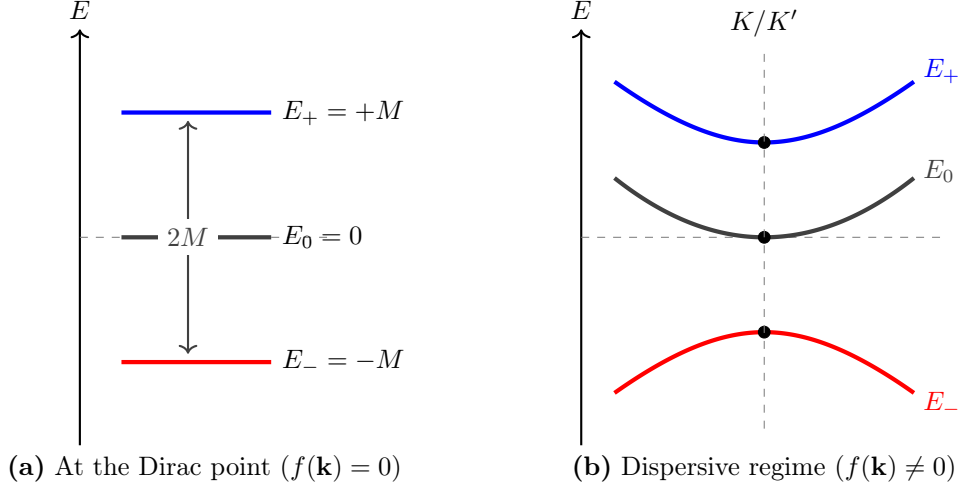

Before applying laser field  ($\Delta_F=0$), the h-BN substrate breaks
the sublattice inversion symmetry of the $\alpha$-$\mathcal{T}_3$
lattice, opening a trivial bandgap of $2M$ at the Dirac points $K$ and $K'$
(Fig.~\ref{fig:static_band_structure}(a)).

To establish the topological baseline we calculate the Berry curvature
$\Omega_n(\mathbf{k})$ for each band $n\in\{-,0,+\}$.
In $2D$ the Berry curvature is the $z$-component of the
curl of the Berry connection
$\mathbf{A}_n = i\langle u_n|\nabla_\mathbf{k}|u_n\rangle$~\cite{berry1984quantal,xiao2010berry}:
\begin{equation}
\Omega_n(\mathbf{k})
= \frac{\partial A_{n,y}}{\partial k_x}
- \frac{\partial A_{n,x}}{\partial k_y}.
\label{eq:berry_curv}
\end{equation}
Expanding $\hat{H}_0(\mathbf{k})$ to first order around
$\mathbf{k}=\xi K+\mathbf{q}$ (valley index $\xi=\pm1$) and applying
second-order perturbation theory, the low-energy Berry curvature of the
valence band is determined by the matrix elements
$\langle n|\partial H/\partial k_j|v\rangle$.
A key structural feature of the $\alpha$-$\mathcal{T}_3$ Hamiltonian
is that there is no direct $A$--$C$ hopping; consequently
$\langle c|\partial H/\partial k_j|v\rangle = 0$ identically, and
only the flat band (hub site $B$) contributes through the
$B$--$C$ coupling $\tau\sin\phi$.
Evaluating the perturbation theory explicitly:
\begin{align}
\langle v|\tfrac{\partial H}{\partial k_x}|f\rangle &= -\tau\sin\phi,
\nonumber\\
\langle v|\tfrac{\partial H}{\partial k_y}|f\rangle &= -i\tau\sin\phi,
\nonumber
\end{align}
Analytical form of the Berry curvature of the valence band takes the form: 
\begin{equation}
\Omega_{-}^{\xi}(\mathbf{q})
= -\xi\,
\frac{\tau^2 M \sin^2\!\phi}
{2\!\left(M^2+\tau^2|\mathbf{q}|^2\right)^{3/2}},
\quad
\sin^2\!\phi = \frac{\alpha^2}{1+\alpha^2}.
\label{eq:berry_valence}
\end{equation}
This result has three important features.
First, the Berry curvature is localized near the Dirac points $K$ and $K'$ and
proportional to both $M$ and $\sin^2\phi$.
Second, it has odd parity under valley exchange,
$\Omega_{-}^{K}(\mathbf{q})=-\Omega_{-}^{K'}(\mathbf{q})$, which is the direct consequence of
time-reversal symmetry.
Third, the prefactor $\sin^2\phi = \alpha^2/(1+\alpha^2)$ increases
monotonically from zero at $\alpha=0$ to $\frac{1}{2}$ at $\alpha=1$
(dice limit), correctly giving a finite Berry curvature at the
dice limit.
Note that Eq.~\eqref{eq:berry_valence} vanishes at $\alpha=0$
(graphene limit), where the $C$ site is completely decoupled
from the $A$--$B$ Dirac sector and carries no Berry curvature of its own.

The Chern number of any band is calculated by integrating over the
full Brillouin zone (BZ)~\cite{thouless1982quantized}:
\begin{equation}
C_n = \frac{1}{2\pi}\int_{\mathrm{BZ}}\Omega_n(\mathbf{k})\,d^2k.
\label{eq:chern_def}
\end{equation}
The time-reversal symmetry of the system without Floquet drive enforces
$\Omega_n^K(\mathbf{q})=-\Omega_n^{K'}(\mathbf{q})$, so the two
valleys cancel exactly upon integration:
\begin{equation}
C_n = \frac{1}{2\pi}\!\int_K\!\Omega_n\,d^2k
    + \frac{1}{2\pi}\!\int_{K'}\!\Omega_n\,d^2k
    = 0.
\label{eq:chern_zero}
\end{equation}
The undriven substrate-supported system is therefore a topologically
trivial insulator ($C=0$) for all bands.
Any non-zero Chern number observed under illumination is a purely
dynamic, Floquet-engineered phenomenon.

\subsection{Floquet--Magnus Derivation and Peierls Substitution}
\label{sec:floquet_magnus}

We introduce an elliptically polarized drive with vector potential
$\mathbf{A}(t)=A_0(\eta\cos\omega t,\,\sin\omega t)$~\cite{seshadri2022engineering}
and incorporate it via the Peierls substitution~\cite{jimenez2012peierls}:
\begin{equation}
 \mathbf{k} \to \mathbf{k}(t)
 = \mathbf{k} + \frac{e}{\hbar}\mathbf{A}(t).
 \label{eq:peierls}
\end{equation}
In the high-frequency limit ($\hbar\omega\gg\tau$) the second order effective time-averaged Hamiltonian is obtained from Floquet-Magnus expansion ~\cite{bukov2015universal,abanin2017effective}:
\begin{equation}
 H_{\mathrm{eff}} \approx H_0 + \frac{[H_{-1},H_{+1}]}{\hbar\omega},
 \label{eq:floquet_magnus}
\end{equation}
where $H_{\pm1}$ are the first Fourier harmonics of $H(t)$.
Evaluating the commutator at the Dirac points gives:
\begin{equation}
 H_{-1}H_{+1} = |V|^2
 \begin{pmatrix} 0&0&0\\ 0&\cos^2\phi&0\\ 0&0&\sin^2\phi \end{pmatrix},
\end{equation}
\begin{equation}
 H_{+1}H_{-1} = |V|^2
 \begin{pmatrix} \cos^2\phi&0&0\\ 0&\sin^2\phi&0\\ 0&0&0 \end{pmatrix}.
\end{equation}
The effective Floquet term $H_{\mathrm{Fl}}=[H_{-1},H_{+1}]/\hbar\omega$
is therefore:
\begin{equation}
 H_{\mathrm{Fl}} =
 \begin{pmatrix}
  -\Delta_F\cos^2\phi & 0 & 0 \\
  0 & \Delta_F\cos 2\phi & 0 \\
  0 & 0 & \Delta_F\sin^2\phi
 \end{pmatrix},
 \label{eq:HFl}
\end{equation}
and the total effective Hamiltonian is:
\begin{gather}
 \hat{H}(\mathbf{k}) = \nonumber \\
 \begin{pmatrix}
  M-\Delta_F\cos^2\phi & -\tau\cos\phi\,f & 0 \\
  -\tau\cos\phi\,f^* & \Delta_F\cos 2\phi & -\tau\sin\phi\,f \\
  0 & -\tau\sin\phi\,f^* & -M+\Delta_F\sin^2\phi
 \end{pmatrix}.
 \label{eq:Heff}
\end{gather}
Setting $M=0$ recovers the massless result of
~\cite{PhysRevB.99.205429}. The alternating signs on the diagonal are due to 
the chirality of the elliptical polarization.

%%%===================================================================
\section{Photo-induced Mass and Gap-closing Conditions}
\label{sec:gaps}
%%%===================================================================

The coupling amplitude $|V|^2$ arising from the Peierls
substitution~\cite{jimenez2012peierls} is:
\begin{equation}
|V|^2 = \left(\frac{eE_0 a\tau}{\hbar\omega}\right)^2
\frac{1+\eta^2}{2},
\label{eq:V2}
\end{equation}
where $E_0=\omega A_0$ is the electric field amplitude, $a$ the
lattice constant, and $\eta$ the ellipticity.
The Floquet mass derived from the second-order Floquet--Magnus
~\cite{bukov2015universal} is:
\begin{equation}
\Delta_F
= \frac{|V|^2}{\hbar\omega}\!\left(\frac{2\eta}{1+\eta^2}\right)
= \frac{e^2E_0^2 a^2\tau^2}{(\hbar\omega)^3}
  \!\left(\frac{2\eta}{1+\eta^2}\right).
\label{eq:DeltaF}
\end{equation}
The value of  $2\eta/(1+\eta^2)$ is maximum at $\eta=1$
(circular polarization) and vanishes for $\eta=0$ (linear), correctly
recovering the absence of a Floquet mass for linearly polarized
light.

For numerical simulations, the $k$-dependent Floquet mass is obtained
by normalizing the imaginary part of the Peierls phase factors to
unity at the $K$ point~\cite{kundu2011tight}:
\begin{equation}
	\begin{split}
		\Delta_F(\mathbf{k}) &= \frac{2\Delta_F}{3\sqrt{3}}
		\Bigl[\sin(\sqrt{3}k_x a) \\
		&\quad - 2\sin\!\left(\frac{\sqrt{3}k_x a}{2}\right)
		\cos\!\left(\frac{3k_y a}{2}\right)\Bigr].
	\end{split}
	\label{eq:DeltaF_k}
\end{equation}

% =====================================================================
%  Section 4.1: Asymmetric Gap-closing Conditions (full rewrite).
%  Explains WHY both valleys must be checked, then treats upper and
%  lower gaps in turn.
% =====================================================================

\subsection{Asymmetric Gap-closing Conditions}
\label{sec:gap_closing}

A gap closes at a Dirac point $\mathbf{k}_D$ when two diagonal elements
of $\hat{H}(\mathbf{k}_D)$ in Eq.~\eqref{eq:Heff} become equal. Because
the elliptical drive breaks time-reversal symmetry, the $k$-dependent
Floquet mass $\Delta_F(\mathbf{k})$ Eq.~\eqref{eq:DeltaF_k} is an
\emph{odd} function of $\mathbf{k}$ and therefore takes opposite signs at
the two inequivalent Dirac points,
\begin{equation}
	\Delta_F(\mathbf{K}) = +\Delta_F,
	\qquad
	\Delta_F(\mathbf{K'}) = -\Delta_F .
	\label{eq:valley_sign}
\end{equation}
The two valleys are thus inequivalent, and the gap-closing conditions
must be evaluated separately at $\mathbf{K}$ and $\mathbf{K'}$. At a Dirac
point the diagonal elements of $\hat{H}(\mathbf{k}_D)$ are $M-\Delta_F(\mathbf{k})\cos^2\phi$
(conduction), $\Delta_F(\mathbf{k})\cos 2\phi$ (flat), and
$-M+\Delta_F(\mathbf{k})\sin^2\phi$ (valence). Only a solution with a
positive (physically reachable) critical intensity is realised.

\subsubsection{Upper gap (conduction--flat band)}
The upper gap closes when the conduction and flat diagonal elements are
equal, $M-\Delta_F(\mathbf{k})\cos^2\phi=\Delta_F(\mathbf{k})\cos 2\phi$.
Evaluating this at both valleys gives $M=\pm\Delta_F(\cos^2\phi+\cos 2\phi)$,
where the upper (lower) sign corresponds to $\mathbf{K}$ ($\mathbf{K'}$).
Since $M>0$ and $\cos^2\phi+\cos 2\phi>0$ for $\alpha<1$, only the
$\mathbf{K}$ valley yields a positive intensity, the $\mathbf{K'}$
solution is negative and hence unphysical. The upper gap therefore has a
\emph{single} closing condition,
\begin{equation}
	\Delta_{F,\mathrm{crit}}^{\mathrm{upper}}(\alpha)
	= \frac{M(1+\alpha^2)}{2-\alpha^2},
	\label{eq:upper_crit}
\end{equation}
which closes at the $\mathbf{K}$ valley. Because its denominator
$2-\alpha^2$ is positive for all $\alpha\in[0,1]$, the upper gap can
always be closed by a finite drive.

\subsubsection{Lower gap (flat--valence band)}
The lower gap closes when the flat and valence diagonal elements are
equal, $\Delta_F(\mathbf{k})\cos 2\phi=-M+\Delta_F(\mathbf{k})\sin^2\phi$.
In contrast to the upper gap, here the two valleys yield \emph{different}
physical conditions. At $\mathbf{K}$ this reads
$\Delta_F\cos 2\phi=-M+\Delta_F\sin^2\phi$, while at $\mathbf{K'}$ the
opposite sign of $\Delta_F(\mathbf{k})$ gives
$-\Delta_F\cos 2\phi=-M-\Delta_F\sin^2\phi$. The two critical intensities
are
\begin{equation}
	\begin{split}
		\Delta_{F,\mathrm{crit}}^{\mathrm{lower},\,\mathbf{K}}(\alpha)
		&= \frac{M(1+\alpha^2)}{2\alpha^2-1}, \\[1ex]
		\Delta_{F,\mathrm{crit}}^{\mathrm{lower},\,\mathbf{K'}}(\alpha)
		&= \frac{M(1+\alpha^2)}{1-2\alpha^2}.
	\end{split}
	\label{eq:lower_valley}
\end{equation}
Only the branch with a positive critical intensity is physical: for
$\alpha<1/\sqrt{2}$ the denominator $2\alpha^2-1$ is negative, so the
$\mathbf{K}$ condition is unphysical and the lower gap closes at
$\mathbf{K'}$, for $\alpha>1/\sqrt{2}$ the roles reverse and the gap
closes at $\mathbf{K}$. The two branches combine into the single
condition
\begin{equation}
	\Delta_{F,\mathrm{crit}}^{\mathrm{lower}}(\alpha)
	= \frac{M(1+\alpha^2)}{\lvert 2\alpha^2-1\rvert},
	\label{eq:lower_crit}
\end{equation}
with the closing occurring at $\mathbf{K'}$ for $\alpha<1/\sqrt{2}$ and at
$\mathbf{K}$ for $\alpha>1/\sqrt{2}$.
\subsection{The $\alpha=1/\sqrt{2}$ Singularity and Phase Stability}
\label{sec:singularity}
Equation~\eqref{eq:lower_crit} diverges when $2\alpha^2-1=0$,
i.e., at $\alpha=1/\sqrt{2}\approx0.707$, where the critical intensity in
both valleys grows without bound. This singularity has a clear physical
meaning: as $\alpha\to1/\sqrt{2}$ the intensity required to close the
lower gap diverges, so the flat--valence inversion is pushed to
arbitrarily strong driving. In this ``protected'' regime the Floquet
drive and the substrate mass act cooperatively to keep the valence band
isolated from the flat band over a wide range of intensities, creating a
robust topological window for the $C=1$ phase
(Fig.~\ref{fig:BandEvolution}(c)). Because the two valleys close the
lower gap at different intensities, the $C=1$ window is bounded from
below by the conduction--flat inversion
Eq.~\eqref{eq:upper_crit} and from above by the earlier of the two
valley closings Eq.~\eqref{eq:lower_crit},
\begin{equation}
	\Delta_{F,\mathrm{crit}}^{\mathrm{upper}}(\alpha)
	\;\le\; \Delta_F \;\le\;
	\Delta_{F,\mathrm{crit}}^{\mathrm{lower}}(\alpha).
	\label{eq:C1_window}
\end{equation}
For the representative case $\alpha=0.5$ this window is
$0.036\,\tau\le\Delta_F\le0.125\,\tau$, within which the band-resolved
Chern numbers are $(C_c,C_f,C_v)=(+1,-1,0)$. For $\alpha>1/\sqrt{2}$ the
denominator of Eq.~\eqref{eq:lower_crit} makes the lower gap closable at
finite intensity in the $\mathbf{K}$ valley, enabling the transition to
the $C=2$ phase.

\begin{figure*}[!t]
    \centering
    \includegraphics[width=0.8\linewidth]{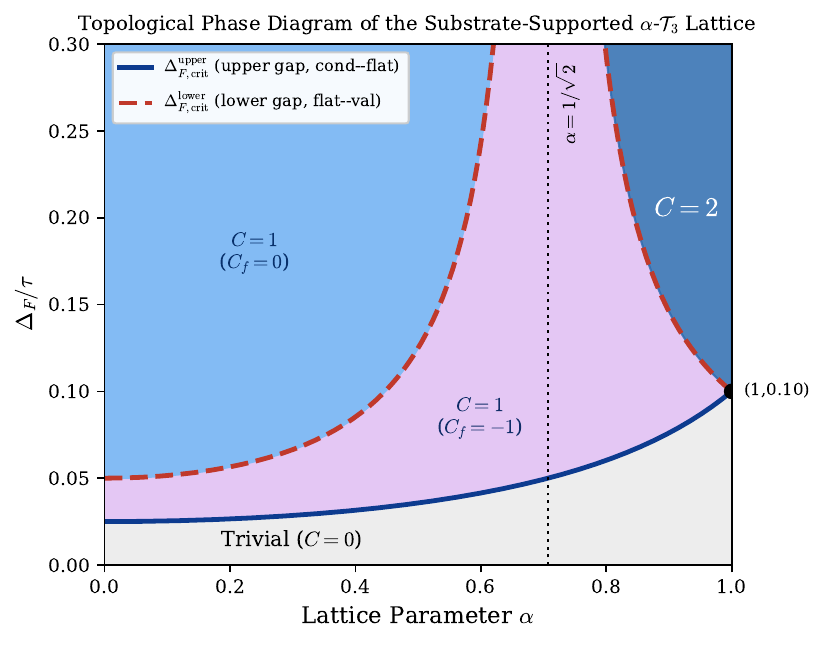}
    	\caption{Topological phase diagram of the $\alpha$-$\mathcal{T}_3$
    		lattice with $M=0.05\,\tau$. The solid blue curve marks the upper
    		gap-closing threshold $\Delta_{F,\mathrm{crit}}^{\mathrm{upper}}$
    		Eq.~\eqref{eq:upper_crit}, where the conduction and flat bands invert,
    		and the dashed red curve marks the valley-resolved lower gap-closing
    		threshold $\Delta_{F,\mathrm{crit}}^{\mathrm{lower}}$
    		Eq.~\eqref{eq:lower_crit}, where the flat and valence bands touch. The
    		two boundaries delimit three topologically distinct regions: a trivial
    		insulator ($C=0$), a Chern insulator with the flat band carrying the
    		non-trivial index ($C=1$, $C_f=-1$) in the window between the two curves,
    		and, once the lower gap has also closed and reopened, a Chern phase in
    		which the flat band returns to $C_f=0$ ($C=1$ for $\alpha<1/\sqrt{2}$,
    		and $C=2$ for $\alpha>1/\sqrt{2}$). The vertical dotted line denotes the
    		geometric singularity at $\alpha=1/\sqrt{2}$, where
    		$\Delta_{F,\mathrm{crit}}^{\mathrm{lower}}$ diverges. The two boundaries
    		converge at the dice limit $\alpha=1$, where they meet at
    		$\Delta_F=2M=0.10\,\tau$.}
    \label{fig:PhaseDiagram}
\end{figure*}

%%%===================================================================
\section{Numerical Results: Band Structure Evolution}
\label{sec:numerics}
%%%===================================================================

All energy scales are normalized to the hopping amplitude $\tau$.
We fix $M=0.05\tau$ throughout the discussion. The choice of $M=0.05\tau$ is reasonable when compared with commensurate
	graphene/h-BN models. With the nearest-neighbor hopping parameter of
	$\tau=1$~eV, this corresponds to a Dirac-point gap of $2M\approx100$~meV.
	Density-functional calculations give substrate-induced gaps ranging from
	about $53$~meV for a lattice-matched structure to $138$~meV for a
	$2\times2$-graphene/$2\times2$-BN supercell, so a Dirac-point gap of
	$2M\approx100$~meV is of the same order as these
	predictions~\cite{giovannetti2007,mdpi2022}.

\subsection{Strategic significance of $\alpha=0.5$}
We focus on $\alpha=0.5$ ($\phi\approx26.57^{\circ}$) as a
representative case in the protected regime ($\alpha<1/\sqrt{2}$). Three
properties make this choice particularly favorable. First, in this
regime the lower gap closes only at the $\mathbf{K'}$ valley
Eq.~\eqref{eq:lower_crit} and at a substantially higher intensity than
the upper gap, so the valence band stays isolated from the flat band
throughout a wide interval of driving strengths. Second, the separation
between the two thresholds is large, providing a wide, experimentally
accessible $C=1$ phase. Third, the system enters $C=1$ already at
$\Delta_F\approx0.036\,\tau$ and remains in it up to
$\Delta_F\approx0.125\,\tau$, so the topological phase can be studied
over an appreciable range of intensity without leaving the window.

\subsection{Calculated phase thresholds}
For $\alpha=0.5$, $M=0.05\,\tau$, the upper threshold is
\begin{equation}
	\Delta_{F,\mathrm{crit}}^{\mathrm{upper}}
	= \frac{0.05\,(1+0.25)}{2-0.25}
	= \frac{0.0625}{1.75}
	\approx 0.036\,\tau ,
	\label{eq:upper_num}
\end{equation}
while the lower gap closes in the $\mathbf{K'}$ valley at
\begin{equation}
	\Delta_{F,\mathrm{crit}}^{\mathrm{lower}}
	= \frac{0.05\,(1+0.25)}{\lvert 2(0.25)-1\rvert}
	= \frac{0.0625}{0.5}
	\approx 0.125\,\tau .
	\label{eq:lower_num}
\end{equation}
The $C=1$ phase therefore occupies the finite window
$0.036\,\tau\le\Delta_F\le0.125\,\tau$. The lower threshold diverges only
as $\alpha\to1/\sqrt{2}$, where the window becomes unbounded.

\subsection{Phase progression}
The key difference from the pristine ($M=0$) lattice is that both
gap-closing thresholds coincide at $\Delta_F=0$ in the massless case. In
the substrate-supported model they are separated by a finite interval
$[\Delta_{F,\mathrm{crit}}^{\mathrm{upper}},
\Delta_{F,\mathrm{crit}}^{\mathrm{lower}}]$, within which only the upper
(conduction--flat) sector has undergone band inversion. This separation
is the origin of the stable $C=1$ topological window.

%%%===================================================================
\section{Results: Phase Transformation}
\label{sec:phases}
%%%===================================================================

\begin{figure*}[!t]
	\centering
	\includegraphics[width=\textwidth]{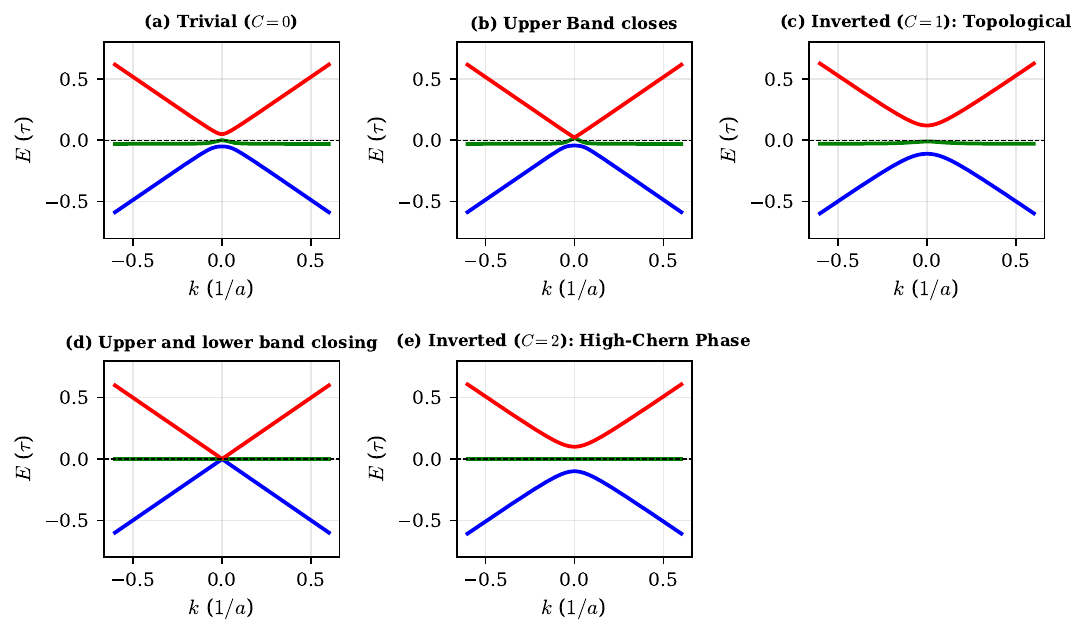}
	\caption{Hierarchical band evolution of the substrate-supported
		$\alpha$-$\mathcal{T}_3$ lattice ($M=0.05\tau$).
		Stages (a)--(c): $\alpha=0.5$, highlighting the finite $C=1$
		topological window bounded by the valley-resolved gap-closing
		thresholds.
		Stages (d)--(e): $\alpha=1$, showing the simultaneous
		gap-closure at the triple-point degeneracy and the subsequent
		$C=2$ phase.}
	\label{fig:BandEvolution}
\end{figure*}

The topological evolution follows a distinct and unique sequential mechanism
(Fig.~\ref{fig:BandEvolution}).

\subsection{Transition dynamics for $\alpha=0.5$}
At $\Delta_F=0$ the h-BN substrate breaks inversion symmetry, creating a
trivial band gap of $2M$ (Fig.~\ref{fig:BandEvolution}(a)). At
$\Delta_F=\Delta_{F,\mathrm{crit}}^{\mathrm{upper}}\approx0.036\,\tau$
the conduction and flat bands touch at the $\mathbf{K'}$ valley, marking
the topological phase boundary (Fig.~\ref{fig:BandEvolution}(b)). For
$\Delta_F>0.036\,\tau$ a topological gap reopens in the upper sector and
the flat band acquires a non-trivial Chern number, while the valence
band stays isolated at the Dirac point, giving the band-resolved indices
$(C_c,C_f,C_v)=(+1,-1,0)$ (Fig.~\ref{fig:BandEvolution}(c)). This $C=1$
phase persists until the lower (flat--valence) gap closes at the
$\mathbf{K'}$ valley at
$\Delta_{F,\mathrm{crit}}^{\mathrm{lower}}\approx0.125\,\tau$
Eq.~\eqref{eq:lower_crit}, so the topological window is the finite
interval $0.036\,\tau\le\Delta_F\le0.125\,\tau$. Beyond this window the
flat--valence inversion redistributes the Berry curvature and the
band-resolved Chern numbers reorganise. All Berry-curvature results for
the $C=1$ phase (Sec.~\ref{sec:berry}) are therefore evaluated at a
representative $\Delta_F=0.08\,\tau$, which lies safely inside the window.

\subsection{Dice limit and high-Chern phase ($\alpha=1$)}

At $\alpha=1$ both critical thresholds converge to
$\Delta_F=2M=0.10\tau$, and all three bands meet at the triple-point
degeneracy (Fig.~\ref{fig:BandEvolution}(d)).
Further increasing $\Delta_F$ reopens both gaps and produces a fully
inverted topological phase with $C=\pm2$
(Fig.~\ref{fig:BandEvolution}(e)).

%%%===================================================================
\section{Berry Curvature Distribution}
\label{sec:berry}
%%%===================================================================
\begin{figure*}[htbp]
		\includegraphics[width=\textwidth]{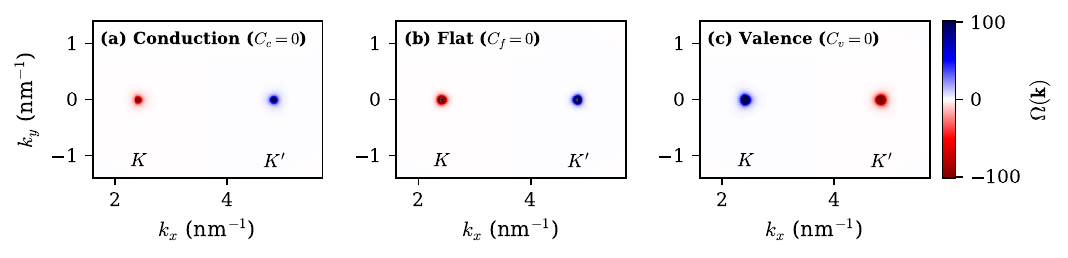}
	
	\caption{Berry curvature $\Omega(\mathbf{k})$ for Case~1
		(trivial insulator, $\Delta_F=0$).
		Equal and opposite peaks at $K$ and $K'$ cancel to yield
		$C_n=0$ for all bands.}
	\label{fig:Case1_Berry}
\end{figure*}

\begin{figure*}[!t]
		\includegraphics[width=\textwidth]{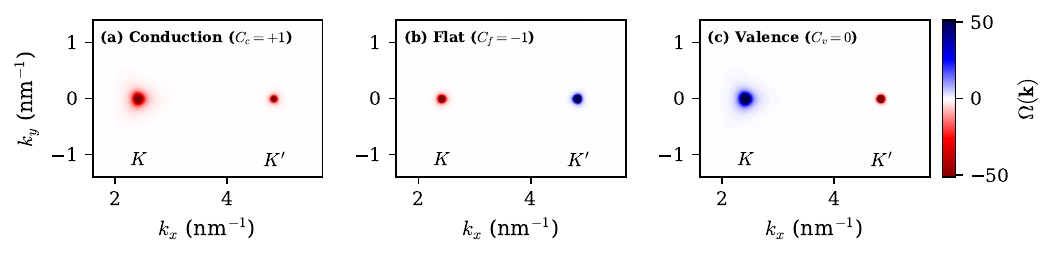}
	\caption{Berry curvature $\Omega(\mathbf{k})$ for Case~2
		($\alpha=0.5$, $\Delta_F\approx0.08\tau$, $C=1$).
		Band inversion is confined to the conduction--flat sector;
		the valence band remains topologically trivial.}
	\label{fig:Case2_Berry}
\end{figure*}

\begin{figure*}[!t]
	\includegraphics[width=\textwidth]{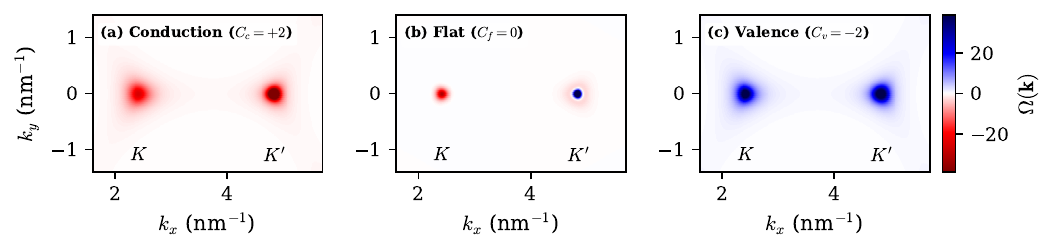}
	\caption{Berry curvature $\Omega(\mathbf{k})$ for Case~3
		(high-Chern phase, $\alpha=0.9$, $\Delta_F=0.40\tau$, $C=2$).
		Double band inversion produces $C_c=+2$ and $C_v=-2$, with the
		flat band returning to a net-zero invariant.}
	\label{fig:Case3_Berry}
\end{figure*}
To verify the topological nature of the photo-induced phases we
calculate the Berry curvature ($\Omega_n(\mathbf{k})$) and Chern number
($C_n$) for each band using the Fukui--Hatsugai--Suzuki (FHS) lattice
method~\cite{gosselin2009berry,nakamura2024chern}. Since the 
drive breaks time-reversal symmetry, the two valleys are inequivalent
and the upper and lower gaps close at different intensities
Eqs.~\eqref{eq:upper_crit} and~\eqref{eq:lower_crit}. We therefore
report the band-resolved Berry curvature within the topological window
$\Delta_{F,\mathrm{crit}}^{\mathrm{upper}}\le\Delta_F\le
\Delta_{F,\mathrm{crit}}^{\mathrm{lower}}$, i.e.\ the regime in which the
conduction--flat inversion has taken place but the flat--valence gap
remains open, so that the flat band carries the non-trivial Chern number.
For $\alpha=0.5$ this corresponds to $\Delta_F=0.08\,\tau$, which lies
safely inside the window $[0.036,0.125]\,\tau$.

\subsection{Case 1: Trivial insulator ($C_c=C_f=C_v=0$)}

In the absence of elliptical drive ($\Delta_F=0$), the Berry curvature forms
equal and opposite dipoles at $K$ and $K'$ within each band
(Fig.~\ref{fig:Case1_Berry}).
These contributions cancel identically upon BZ integration, giving
$C_n=0$ for all bands, consistent with Eq.~\eqref{eq:chern_zero}.

\subsection{Case 2: Chern insulator ($C_c=+1,\,C_f=-1,\,C_v=0$)}

Once $\Delta_F\gtrsim0.036\tau$, the conduction and flat bands
undergo topological inversion (Fig.~\ref{fig:Case2_Berry}).
The conduction band ($C_c=+1$) develops asymmetric positive peaks that
dominate, reflecting the net Berry flux of $+1$.
Corresponding negative peaks appear in the flat band ($C_f=-1$),
conserving the total Chern number $C_c+C_f+C_v=0$.
The valence band ($C_v=0$) remains balanced and decoupled, protected
by the geometric singularity.

\subsection{Case 3: High-Chern phase ($C_c=+2,\,C_f=0,\,C_v=-2$)}

For $\alpha>1/\sqrt{2}$ and sufficiently large $\Delta_F$
(Fig.~\ref{fig:Case3_Berry}), sharp high-magnitude peaks in the
conduction band ($C_c=+2$) produce a total Berry flux of $+2$.
The flat band ($C_f=0$) returns to a balanced configuration and its
net Berry flux vanishes.
In high-$\alpha$ geometries the lower-gap protection is overcome,
inducing a high-Chern response in the valence band ($C_v=-2$).
The total Chern number is conserved: $C_c+C_f+C_v=0$, consistent with
the TKNN sum rule~\cite{hatsugai1993chern}.

The band-resolved Chern numbers for all three phases are
	summarized in Table~\ref{tab:chern}.

\begin{table}[htbp]
	\centering
	\begin{tabular}{lcccc}
		\hline\hline
		Phase & $C_c$ & $C_f$ & $C_v$ & $C$ \\
		\hline
		Trivial insulator & $0$  & $0$  & $0$  & $0$ \\
		Chern insulator   & $+1$ & $-1$ & $0$  & $1$ \\
		High-Chern phase  & $+2$ & $0$  & $-2$ & $2$ \\
		\hline\hline
	\end{tabular}
	\caption{Band-resolved Chern numbers of the conduction ($C_c$), flat
		($C_f$), and valence ($C_v$) bands in the three topological phases. In
		each phase the band Chern numbers sum to zero, and the total Chern number
		of the filled bands gives the quantized Hall response $C$. The values for
		the Chern-insulator phase are those of the topological window
		$\Delta_{F,\mathrm{crit}}^{\mathrm{upper}}\le\Delta_F\le
		\Delta_{F,\mathrm{crit}}^{\mathrm{lower}}$, where the conduction--flat
		inversion has occurred but the flat--valence gap remains open, so that
		the flat band carries the non-trivial index $C_f=-1$.}
	\label{tab:chern}
\end{table}

\section{Bulk--Boundary Correspondence: Chiral Edge States}
\label{sec:edge}

\begin{figure*}[!t]
	\centering
	\includegraphics[width=\linewidth]{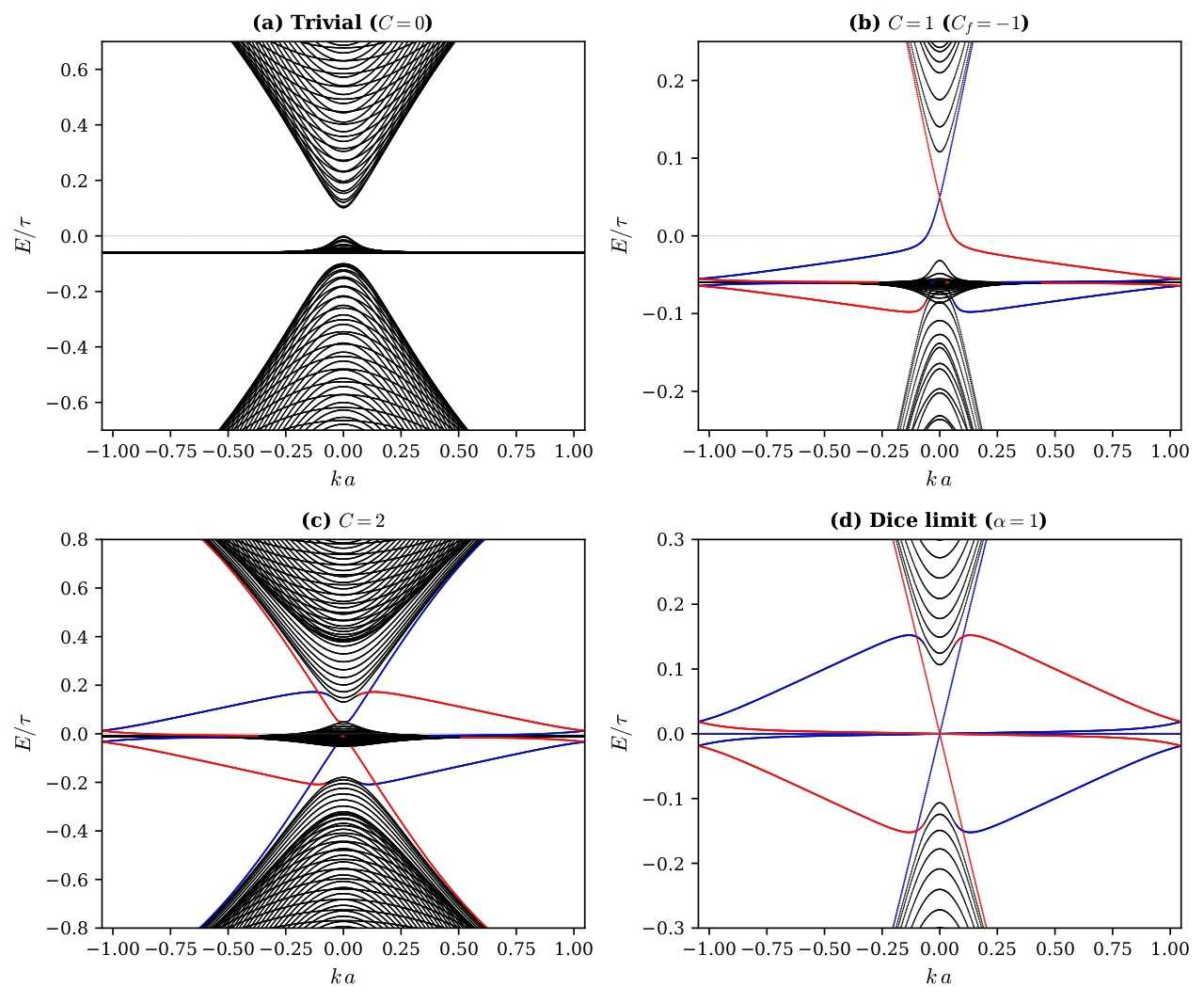}
	\caption{Chiral edge-state spectra of an armchair
		$\alpha$-$\mathcal{T}_3$ nanoribbon ($M=0.10\,\tau$, used only to
		widen the gaps for clarity; the topology is unchanged at
		$M=0.05\,\tau$). Red/blue: opposite-edge states; black: bulk.
		(a)~Trivial ($C=0$), (b)~$C=1$ with $C_f=-1$, (c)~$C=2$,
		(d)~dice limit ($\alpha=1$).}
	\label{fig:edge}
\end{figure*}

Having established the band-resolved Chern numbers in the previous section, we now validate these bulk topological invariants through the bulk--boundary correspondence. According to this principle, a bulk band with a non-zero Chern number must be accompanied by chiral edge states traversing the corresponding bulk energy gap. To verify this correspondence, we compute the energy spectrum of an armchair $\alpha$-$\mathcal{T}_3$ nanoribbon in the parameter regime where the Floquet-engineered flat band acquires a non-trivial Chern number. The ribbon is periodic along one direction (Bloch momentum $k$) and
finite in the transverse direction, so that both Dirac valleys $K$ and
$K'$ project onto $k=0$. States are coloured by their spatial
localization: red and blue denote modes bound to opposite edges of the
ribbon, while the bulk continuum is shown in black. Figure~\ref{fig:edge}
presents the edge spectra for the four characteristic regimes of the
phase diagram.

Within the $C=1$ window the bulk gaps are intrinsically small (of order a
few $\times 10^{-2}\,\tau$ for $M=0.05\,\tau$), which obscures the edge
modes in a ribbon calculation. To resolve the edge spectra clearly we
therefore set $M=0.10\,\tau$ in Fig.~\ref{fig:edge}; this widens the bulk
gaps without altering the band topology. We have verified numerically,
using the Fukui--Hatsugai--Suzuki method, that the band-resolved Chern
numbers are identical for $M=0.05\,\tau$ and $M=0.10\,\tau$ throughout
each phase, so the edge-state structure shown here applies unchanged to
the physical case $M=0.05\,\tau$ discussed in the rest of this work.

In the absence of the Floquet drive the substrate opens a trivial gap of
width $2M$ around the flat band. The spectrum is fully gapped and no
in-gap states connect the bulk bands: the ribbon is an ordinary
insulator, consistent with $C_c=C_f=C_v=0$. 
Once the drive exceeds the upper critical intensity
$\Delta_{F,\mathrm{crit}}^{\mathrm{upper}}$, the conduction and flat
bands invert and the flat band acquires a Chern number $C_f=-1$, while
the valence band remains decoupled ($C_v=0$). The edge spectrum
[Fig.~\ref{fig:edge}(b)] shows exactly one chiral branch per edge that
disperses continuously across the flat--conduction gap, connecting the
flat band to the conduction band as $k$ sweeps through the zone centre.
This single pair of counter-propagating edge modes is the boundary
signature of $|C_f|=1$.

We emphasize that this section addresses only the regime in which the
flat band carries the topological index. It is instructive to note the
role of the small residual gap. Because the flat and valence bands lie
close in energy, the chiral edge branch dips slightly below $E=0$ near
the zone centre; however, it never merges with the valence continuum. In
other words, the edge mode passes below zero energy but does not
\emph{connect} to the valence band---its two ends terminate on the flat
and conduction bands. The topological content is encoded in this
connectivity, not in the energy at which the crossing occurs: the mode
links the flat and conduction sectors, confirming $C_f=-1$ and leaving
the valence band topologically inert.

For $\alpha>1/\sqrt{2}$ and sufficiently strong driving the lower
(flat--valence) gap also closes and reopens, so that both the conduction
and valence bands invert through the flat band. The edge spectrum now
exhibits \emph{two} co-propagating chiral branches per edge crossing the
gap, the hallmark of $|C|=2$, with the band-resolved indices
$(C_c,C_f,C_v)=(+2,0,-2)$. Here the flat band has returned to a trivial
configuration and the topological charge is shared between the
conduction and valence bands.
At the dice point the upper and lower gap-closing thresholds coincide at
$\Delta_F=2M$, and all three bands touch simultaneously at the triple
point before reopening. The resulting edge spectrum displays the
two-chiral-mode structure of the $C=2$ phase, now emerging symmetrically
about $E=0$ as a consequence of the restored electron--hole symmetry of
the dice lattice.

Taken together, panels (a)--(d) trace the complete topological evolution
driven by the elliptical field. As the drive is increased from zero, the
conduction--flat inversion first transfers a unit of Berry flux to the
flat band ($C_f:0\to-1$), producing a single chiral edge mode. Upon
further increasing the drive, the valence band touches the flat band at
$K'$; this second inversion returns the flat band to a trivial state
($C_f\to 0$) and redistributes the topological charge between the
conduction and valence bands, doubling the number of chiral edge modes.
The nanoribbon spectra thus provide a direct, real-space confirmation of
the valley-resolved gap-closing sequence and of the band-resolved Chern
numbers obtained from the bulk Berry curvature.

\subsection{Advantages of elliptical polarization}

The advantage of elliptical light is clear from the phase diagram
(Fig.~\ref{fig:PhaseDiagram}), whose vertical axis is the Floquet mass
$\Delta_F$ given in Eq.~\eqref{eq:DeltaF}. The value of $\Delta_F$ can
be raised in two ways, either by increasing the laser power $E_0$ or by
increasing the ellipticity $\eta$ toward circular light. With circular
light the ellipticity is fixed at $\eta=1$, so the only way to move
across the phase diagram is to change the laser power itself. Elliptical
light adds a second control, since the same phase boundaries can be
crossed at a fixed laser power simply by tuning the polarization $\eta$.
This gives a continuous and convenient way to scan the $C=0$, $C=1$ and
$C=2$ phases in our system without repeatedly adjusting the beam
intensity~\cite{seshadri2022engineering}.

% =====================================================================
%  SECTION: Valley-Resolved Hall Transport Dynamics
%  Simple-English section with Chern table, per-case numerical
%  temperature analysis, and an experimental summary.
% =====================================================================

\section{Valley-Resolved Hall Transport Dynamics}
\label{sec:hall}

Having established the band-resolved Chern numbers, we now study how they
appear in the anomalous Hall conductivity $\sigma_{xy}$ and how robust the
resulting plateaus are against temperature. The Hall conductivity follows
from the TKNN sum rule~\cite{nagaosa2010anomalous,kubo1957statistical},
\begin{equation}
	\sigma_{xy}(\mu) = \frac{e^2}{h}\sum_{n:\,E_n<\mu} C_n ,
	\label{eq:tknn}
\end{equation}
where the sum runs over every band whose energy lies below the Fermi
level $\mu$. Because the three bands carry different Chern numbers, the
Hall conductivity forms a staircase as $\mu$ is swept across the gaps. The
key point is that the \emph{height} and the \emph{position} of each
plateau tell us directly which bands carry the topological charge.

Table~\ref{tab:chern} lists the band-resolved Chern numbers in the three
regions of the phase diagram (Fig.~\ref{fig:PhaseDiagram}). We now examine the
Hall response and its temperature dependence in each region. Throughout,
we take the hopping amplitude $\tau=1$~eV, so that the thermal energy is
$k_BT \approx 8.6$, $17.2$, and $25.9$~meV at $T=100$, $200$, and
$300$~K, respectively.

\begin{table}[htbp]
	\centering
	\begin{tabular}{lcccc}
		\hline\hline
		Phase & $C_c$ & $C_f$ & $C_v$ & $C$ \\
		\hline
		Trivial insulator      & $0$  & $0$  & $0$  & $0$ \\
		Chern insulator (flat) & $+1$ & $-1$ & $0$  & $1$ \\
		Chern insulator (valence) & $+1$ & $0$  & $-1$ & $1$ \\
		High-Chern phase       & $+2$ & $0$  & $-2$ & $2$ \\
		\hline\hline
	\end{tabular}
		\caption{Band-resolved Chern numbers of the conduction ($C_c$), flat
		($C_f$), and valence ($C_v$) bands in the three regions of the phase
		diagram (Fig.~\ref{fig:PhaseDiagram}). The band Chern numbers sum to zero in
		each region, and the total Chern number of the filled bands gives the
		quantized Hall response $C$.}
	\label{tab:chern}
\end{table}

\subsection{Case 1: conduction and flat bands share the charge ($C=1$,
	$C_f=-1$)}
Inside the topological window ($\alpha=0.5$, $\Delta_F=0.06\,\tau$) the
conduction and flat bands share the Chern number, $(C_c,C_f,C_v)=
(+1,-1,0)$. Sweeping the Fermi level gives the sequence
$0 \to 0 \to e^2/h \to 0$: the quantized plateau appears only once the
Fermi level lies above the flat band. In other words, one must fill the
flat band to switch on the Hall response. This is the direct transport
fingerprint that the flat band itself carries the topological charge.

\begin{figure*}[!t]
	\centering
	\includegraphics[width=\linewidth]{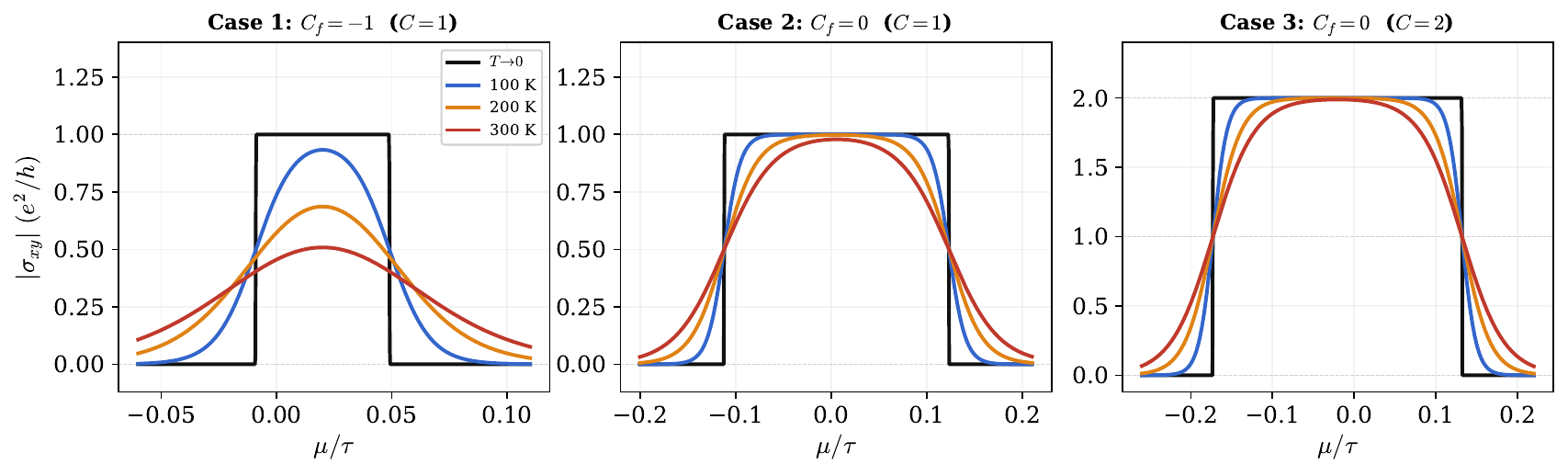}
	\caption{Anomalous Hall conductivity $|\sigma_{xy}|$ as a function of
		Fermi level $\mu$ at temperatures $T\to0$, $100$, $200$, and $300$~K
		($\tau=1$~eV). Each panel shows a quantized plateau bounded by two
		band edges; thermal broadening rounds both edges, so a narrow plateau
		erodes with temperature while a wide one survives.
		\textbf{(a)}~Case~1, $(C_c,C_f,C_v)=(+1,-1,0)$: the $e^2/h$ plateau
		sits in the narrow flat--conduction gap ($\approx 58$~meV) and falls
		to $\approx 51\%$ of its quantized value at $300$~K.
		\textbf{(b)}~Case~2, $(+1,0,-1)$: the $e^2/h$ plateau spans a wide
		window ($\approx 235$~meV) and remains $\approx 98\%$ quantized at
		$300$~K.
		\textbf{(c)}~Case~3, $(+2,0,-2)$: the $2e^2/h$ plateau spans
		$\approx 305$~meV and stays within $1\%$ of quantization at
		$300$~K. The narrow-gap Case~1 is the most fragile, which is why its
		finite-temperature curves do not share a single crossing point,
		unlike the wide-gap Cases~2 and~3.}
	\label{fig:hall}
\end{figure*}

The plateau is protected by the flat--conduction gap, which is only
$\approx 58$~meV wide because the flat band lies close in energy to the
conduction band. As shown in Fig.~\ref{fig:hall}(a), this narrow gap makes
the plateau fragile. The quantized value survives fully at $T=0$, but
drops to about $93\%$ at $100$~K, to $69\%$ at $200$~K, and to only
$51\%$ at $300$~K. At room temperature the plateau is therefore barely
half of its ideal value.

\subsection{Case 2: conduction and valence bands share the charge ($C=1$,
	$C_f=0$)}
For stronger driving ($\alpha=0.5$, $\Delta_F=0.20\,\tau$) the
flat--valence gap has closed and reopened, transferring the flat band's
charge to the valence band. The Chern numbers are now
$(C_c,C_f,C_v)=(+1,0,-1)$: the conduction and valence bands share the
charge while the flat band is topologically inert. The Hall sequence is
$0 \to e^2/h \to e^2/h \to 0$, and because the flat band adds nothing, the
plateau extends across both gaps, giving an effective quantized window
about $235$~meV wide.

This wide window makes the plateau very robust Fig.~\ref{fig:hall}(b).
The quantized value stays at essentially $100\%$ up to $200$~K and is
still about $98\%$ at $300$~K. The same integer Hall response ($C=1$) is
therefore far more stable in Case~2 than in Case~1, simply because the
charge sits on well-separated bands.

\subsection{Case 3: high-Chern phase ($C=2$, $C_f=0$)}
For $\alpha>1/\sqrt{2}$ ($\alpha=0.9$, $\Delta_F=0.40\,\tau$) a double
inversion produces $(C_c,C_f,C_v)=(+2,0,-2)$. The flat band is again
trivial and the charge is shared between the conduction and valence
bands. The Hall sequence is $0 \to 2e^2/h \to 2e^2/h \to 0$, with the
$2e^2/h$ plateau spanning an effective window of about $305$~meV.

This is the widest window of the three, so the $C=2$ plateau is the most
robust Fig.~\ref{fig:hall}(c). It remains within $1\%$ of $2e^2/h$ even
at $300$~K. A high-Chern response, often expected to be delicate, is here
protected by a large gap and survives to room temperature.

\subsection{Which case is best for experiment?}
The three cases show that the same quantized Hall response can be either
fragile or robust depending on \emph{which} bands carry the Chern number.
The controlling factor is the size of the gap in which the Fermi level
sits, compared with the thermal energy $k_BT$:
\begin{itemize}
	\item \textbf{Case 1} ($C_f=-1$): gap $\approx 58$~meV; plateau falls to
	$\approx 51\%$ at $300$~K. This case is the most fragile, but it is also
	the most \emph{interesting}, because the Hall plateau switches on only
	when the flat band is filled---a clean experimental signature that the
	flat band is topological. It is best observed at low temperature
	($\lesssim 100$~K).
	\item \textbf{Case 2} ($C_f=0$, $C=1$): gap $\approx 235$~meV; plateau
	$\approx 98\%$ at $300$~K. This is the most robust way to obtain the
	$e^2/h$ plateau and is well suited to room-temperature measurement.
	\item \textbf{Case 3} ($C_f=0$, $C=2$): gap $\approx 305$~meV; plateau
	$\approx 99\%$ at $300$~K. This gives the largest Hall signal
	($2e^2/h$) and is also robust to room temperature.
\end{itemize}
For an experiment that simply wants a clean, temperature-stable quantized
plateau, Cases~2 and~3 are the best choices, with Case~3 offering the
largest signal. For an experiment that aims to \emph{demonstrate the
	topological nature of the flat band}, Case~1 is the target, but it
requires low temperatures because of its narrow protecting gap. In short,
the flat-band topological plateau is the most scientifically striking but
the most thermally delicate, while transferring the charge to the
conduction and valence bands (Cases~2 and~3) trades that striking
signature for far greater robustness.

% =====================================================================
%  Figure caption for the Hall conductivity figure (Fig. \ref{fig:hall})
% =====================================================================

\section{Anomalous Thermoelectric Transport}
\label{seed}

To provide further experimentally verifiable signatures, we evaluate
the thermoelectric transport coefficients using the multi-band
semiclassical Boltzmann formulation combined with the Mott relation~\cite{mahan2000many}.
In the low-temperature limit ($k_BT\ll\tau$) the transport integrals are:

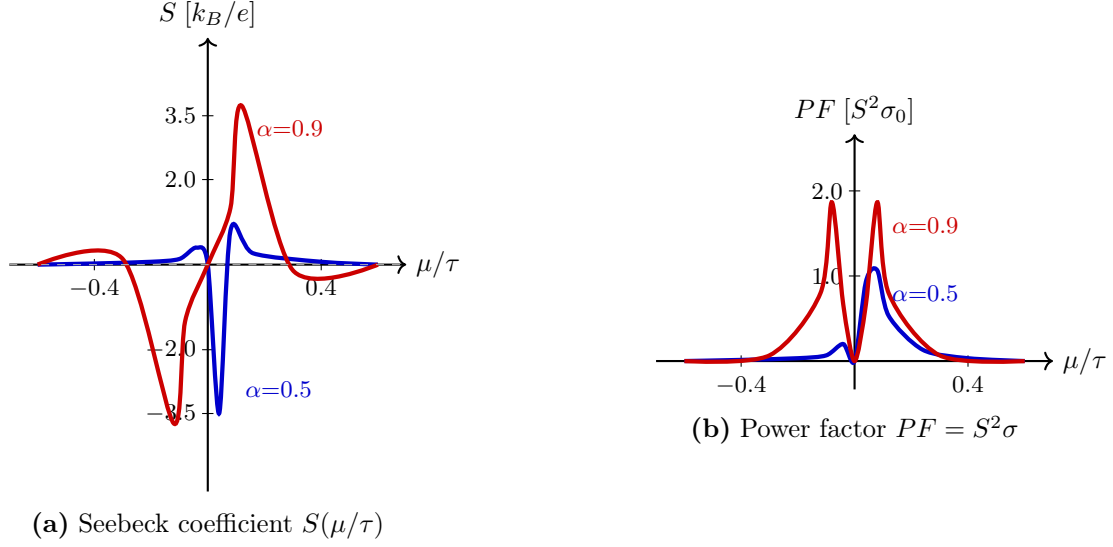
\begin{figure*}[!t]
	\centering
	\begin{minipage}{0.48\textwidth}
		\centering
		\begin{tikzpicture}[scale=0.75]
			\draw[thick,->] (-3.5,0) -- (3.5,0) node[right] {$\mu/\tau$};
			\draw[thick,->] (0,-4.0) -- (0,4.0) node[above] {$S\;[k_B/e]$};
			\draw[thin] (-0.1,1.5) -- (0.1,1.5) node[left, xshift=-3pt] {\small $2.0$};
			\draw[thin] (-0.1,-1.5) -- (0.1,-1.5) node[left, xshift=-3pt] {\small $-2.0$};
			\draw[thin] (-0.1,2.625) -- (0.1,2.625) node[left, xshift=-3pt] {\small $3.5$};
			\draw[thin] (-0.1,-2.625) -- (0.1,-2.625) node[left, xshift=-3pt] {\small $-3.5$};
			\draw[thin] (2.0,0.1) -- (2.0,-0.1) node[below] {\small $0.4$};
			\draw[thin] (-2.0,0.1) -- (-2.0,-0.1) node[below] {\small $-0.4$};
			\draw[ultra thick, blue!80!black, smooth] plot coordinates {
				(-3.0,0) (-2.0,0.02) (-1.0,0.05) (-0.5,0.1) (-0.2,0.3)
				(0.0,0.0) (0.2,-2.625) (0.4,0.6) (0.8,0.2) (2.0,0.05) (3.0,0.0)
			};
			\draw[ultra thick, red!80!black, smooth] plot coordinates {
				(-3.0,0) (-1.5,0.1) (-0.6,-2.8) (-0.4,-1.0)
				(0.0,0.0)
				(0.4,1.0) (0.6,2.8) (1.5,-0.1) (3.0,0.0)
			};
			\draw[dashed, gray!60] (-3.5,0) -- (3.5,0);
			\node[blue!80!black, right] at (0.5,-2.2) {\small $\alpha{=}0.5$};
			\node[red!80!black, right]  at (0.7, 2.4) {\small $\alpha{=}0.9$};
			\node[below] at (0,-4.2) {\textbf{(a)} Seebeck coefficient $S(\mu/\tau)$};
		\end{tikzpicture}
	\end{minipage}
	\hfill
	\begin{minipage}{0.48\textwidth}
		\centering
		\begin{tikzpicture}[scale=0.75]
			\draw[thick,->] (-3.5,0) -- (3.5,0) node[right] {$\mu/\tau$};
			\draw[thick,->] (0,-0.5) -- (0,4.0) node[above] {$PF\;[S^2\sigma_0]$};
			\draw[thin] (-0.1,1.5) -- (0.1,1.5) node[left, xshift=-3pt] {\small $1.0$};
			\draw[thin] (-0.1,3.0) -- (0.1,3.0) node[left, xshift=-3pt] {\small $2.0$};
			\draw[thin] (2.0,0.1) -- (2.0,-0.1) node[below] {\small $0.4$};
			\draw[thin] (-2.0,0.1) -- (-2.0,-0.1) node[below] {\small $-0.4$};
			\draw[ultra thick, blue!80!black, smooth] plot coordinates {
				(-3.0,0) (-2.0,0.02) (-1.0,0.05) (-0.5,0.1) (-0.2,0.3)
				(0.0,0.0) (0.2,1.4) (0.4,1.6) (0.6,0.8) (1.2,0.2) (2.0,0.05) (3.0,0.0)
			};
			\draw[ultra thick, red!80!black, smooth] plot coordinates {
				(-3.0,0) (-1.5,0.1) (-0.6,1.2) (-0.4,2.8) (-0.2,1.0)
				(0.0,0.0)
				(0.2,1.0) (0.4,2.8) (0.6,1.2) (1.5,0.1) (3.0,0.0)
			};
			\node[blue!80!black, right] at (0.5, 1.2) {\small $\alpha{=}0.5$};
			\node[red!80!black, right]  at (0.5, 2.4) {\small $\alpha{=}0.9$};
			\node[below] at (0,-0.8) {\textbf{(b)} Power factor $PF=S^2\sigma$};
		\end{tikzpicture}
	\end{minipage}
	\caption{Floquet-engineered thermoelectric transport at $k_BT=0.02\tau$.
		(a) Seebeck coefficient $S(\mu/\tau)$: the protected geometry
		($\alpha=0.5$, blue) shows a highly asymmetric single peak
		($S\approx-3.5\,k_B/e$) at the upper band edge, while the
		high-Chern regime ($\alpha=0.9$, red) exhibits a symmetric
		double-peak structure ($S\approx\pm4.2\,k_B/e$) reflecting both
		gap inversions.
		(b) Power factor $PF=S^2\sigma$: the high-Chern state produces
		a doubled peak due to simultaneous enhancement of both $S$ and
		$\sigma$ near the resonance zones.}
	\label{fig:thermoelectric_properties}
\end{figure*}

\begin{equation}
L_m(\mu) = \int_{-\infty}^{\infty}
\!\left(-\frac{\partial f_0}{\partial E}\right)
(E-\mu)^m \Sigma(E)\,dE,
\label{eq:Lm}
\end{equation}
where $f_0(E,\mu,T)=[1+e^{(E-\mu)/k_BT}]^{-1}$ is the Fermi--Dirac
distribution and $\Sigma(E)$ is the energy-dependent transport
distribution function.
The electrical conductivity is $\sigma(\mu)=e^2 L_0(\mu)$ and the
Seebeck coefficient is:
\begin{equation}
S(\mu) = -\frac{1}{eT}\frac{L_1(\mu)}{L_0(\mu)}
\approx -\frac{\pi^2k_B^2T}{3e}
\frac{d\ln\sigma(E)}{dE}\bigg|_{E=\mu}.
\label{eq:seebeck}
\end{equation}

For the protected geometry ($\alpha=0.5$), the Seebeck response
exhibits a sharp, highly asymmetric bipolar peak at the upper band
edge.
As $\mu$ moves from the flat band toward the conduction band edge
$E\approx M-\Delta_F\cos^2\phi$, the steep slope of the density of
states yields a peak $S\approx-3.5\,k_B/e$.
Within the topological window the valence band stays gapped from the flat
band, producing a flat thermoelectric profile for $\mu<0$.
This unidirectional response is a direct transport signature of the
protected single-gap topological window.

In the high-Chern regime ($\alpha=0.9$), both energy gaps undergo
simultaneous inversion, making the transport distribution sensitive
at both band edges.
This yields symmetric dual Seebeck peaks ($S\approx\pm4.2\,k_B/e$)
and a doubled power factor, confirming that the high-Chern Floquet
state provides a superior platform for tunable low-temperature
thermoelectric energy conversion.

%%%===================================================================
\section{Conclusion}
\label{sec:conclusions}
%%%===================================================================

We investigated the topological phase transitions of the
$\alpha$-$\mathcal{T}_3$ lattice under the combined influence of an
h-BN substrate and off-resonant elliptical Floquet driving.
With a staggered potential $M=0.05\tau$, we identified a unique
topological protection mechanism that fundamentally distinguishes the
substrate-supported system from the pristine lattice.
Our main findings are as follows.

First, we identified a fundamental geometric singularity at
$\alpha=1/\sqrt{2}$, independent of $M$, where the lower gap-closing
threshold diverges. Our analytical derivation further revealed that,
because the elliptical drive breaks time-reversal symmetry, the lower
(flat--valence) gap closes at \emph{different} intensities in the two
valleys. The $k$-dependent Floquet mass is odd in $\mathbf{k}$, so the
lower-gap threshold takes the valley-resolved form
\begin{equation}
	\Delta_{F,\mathrm{crit}}^{\mathrm{lower}}(\alpha)
	= \frac{M(1+\alpha^2)}{\lvert 2\alpha^2-1\rvert},
\end{equation}
with the gap closing at the $\mathbf{K'}$ valley for $\alpha<1/\sqrt{2}$
and at the $\mathbf{K}$ valley for $\alpha>1/\sqrt{2}$. Together with the
upper-gap threshold
$\Delta_{F,\mathrm{crit}}^{\mathrm{upper}}=M(1+\alpha^2)/(2-\alpha^2)$,
this bounds a finite $C=1$ topological window in which the flat band
carries the non-trivial Chern number. The two thresholds coincide only at
$\alpha=1/\sqrt{2}$, where the window closes and the lower-gap threshold
diverges.

Second, as $\alpha\to1$ the system symmetry is restored and the two
critical intensities converge, enabling simultaneous double-gap
closure, resulting in the  $C=2$ phase.

Third, the ellipticity provides a convenient experimental knob~\cite{seshadri2022engineering}. Since the Floquet mass scales as $\Delta_F\propto 2\eta/(1+\eta^2)$, tuning the polarization $\eta$ at fixed laser power moves the system vertically across the phase diagram, allowing the $C=0$, $C=1$, and $C=2$ phases to be accessed without changing the beam intensity.

The predicted topological phases can be detected experimentally 
through transport measurements. 
The Hall conductivity shows quantized steps at 
$\sigma_{xy} = e^2/h$ for $C=1$ and 
$\sigma_{xy} = 2e^2/h$ for $C=2$, where each valley contributes 
equally with $\frac{1}{2}e^2/h$~\cite{nagaosa2010anomalous,chang2013experimental}. The two phases differ noticeably in thermal robustness.The robustness of each quantized plateau is set by the size of the gap in
which the Fermi level lies: the flat-band $C=1$ plateau, protected by the
narrow flat--conduction gap ($\approx 58$~meV), is the most fragile and is
best observed below $\sim100$~K, whereas the $C=2$ plateau, protected by a
much wider gap, remains robust up to room temperature.
The Seebeck coefficient shows a single sharp peak in the protected 
regime ($\alpha=0.5$) and two symmetric peaks in the high-Chern 
regime ($\alpha=0.9$), serving as a direct experimental signature 
of the underlying topological phase.

Finally, our results provide a necessary correction to prior massless
models~\cite{PhysRevB.99.205429}. The introduction of a staggered
potential $M$ completely rewrites the phase diagram.
The initial state is a trivial insulator ($C=0$); the transition to
$C=1$ proceeds through a controlled band-crossing involving only the
upper gap; and the transition to $C=2$ requires both finite $\alpha>1/\sqrt{2}$
and sufficient drive intensity.
This provides a more realistic framework for experimentalists seeking
to realize stable high-Chern-number phases in h-BN-supported
$\alpha$-$\mathcal{T}_3$ materials.

\clearpage
\onecolumn
\setlength{\bibsep}{6pt}
\bibliographystyle{elsarticle-num}
\bibliography{References}

\end{document}